\documentclass[%
 preprint,
 amsmath,amssymb,
 aps,
prd,
showkeys,
showpacs,
]{revtex4-1}

\usepackage[dvipsnames]{xcolor}
\usepackage[normalem, normalbf]{ulem}
\usepackage{graphicx}
\usepackage{dcolumn}
\usepackage{bm}
\usepackage{amsmath, amsthm,amssymb}
\usepackage{changes}
\usepackage{epstopdf}

\newcommand*{\affmark}[1][*]{\textsuperscript{#1}}

\begin{document}

\title{Enhanced detection of high frequency gravitational waves using optically diluted optomechanical filters}

\email[Correspondence: ]{mpagephys@gmail.com}
\author{Michael Page\affmark[1], Jiayi Qin\affmark[1,2], James La Fontaine\affmark[1], Chunnong Zhao\affmark[1], Li Ju\affmark[1], David Blair\affmark[1]}
\affiliation{\affmark[1]OzGrav-UWA, University of Western Australia, 35 Stirling Highway, Crawley, Western Australia, 6009, Australia}
\affiliation{\affmark[2]Centre for Quantum Computation and Communication Technology, Research School of Physics and Engineering, Australian National University, Canberra, ACT 2601, Australia}

\vspace{2pc}

\begin{abstract}

Detections of gravitational waves (GW) in the frequency band 35 Hz to 500 Hz have led to the birth of GW astronomy. Expected signals above 500 Hz, such as the quasinormal modes of lower mass black holes and neutron star mergers signatures are currently not detectable due to increasing quantum shot noise at high frequencies. Squeezed vacuum injection has been shown to allow broadband sensitivity improvement, but this technique does not change the slope of the noise at high frequency. It has been shown that white light signal recycling using negative dispersion optomechanical filter cavities with strong optical dilution for thermal noise suppression can in principle allow broadband high frequency sensitivity improvement. Here we present detailed modelling of AlGaAs/GaAs optomechanical filters to identify the available parameter space in which such filters can achieve the low thermal noise required to allow useful sensitivity improvement at high frequency. Material losses, the resolved sideband condition and internal acoustic modes dictate the need for resonators substantially smaller than previously suggested. We identify suitable resonator dimensions and show that a 30 $\mu$m scale cat-flap resonator combined with optical squeezing allows 8 fold improvement of strain sensitivity at 2 kHz compared with Advanced LIGO. This corresponds to a detection volume increase of a factor of 500 for sources in this frequency range.

\end{abstract}

\maketitle

\section{Introduction}

The detection of gravitational waves (GW) by Advanced LIGO and VIRGO \cite{GWDiscovery, GWDiscovery2, GWDiscovery3, GWThreeDetector, NSDiscovery} confirmed the existence and detectability of GW and opened the field of GW astronomy. The discoveries showed the inspiral of black holes and one neutron star binary, and in one case, the quasi-normal mode of the final black hole with mass 62 M$_{\odot}$ was identified \cite{GWDiscovery, DiscoveryCompanion}. The electromagnetic outburst from the neutron star binary merger allowed the speed of GW to be measured \cite{NSDiscovery}.

Predicted waveforms for neutron star coalescence show strong structure in the frequency range 1-4 kHz \cite{ClarkQNM, MaioneEOS}. Observations in this band would reveal both the equation of state of the neutron stars, and the dynamics of collapse to a black hole. The ringdown characteristics of a black hole quasi-normal mode are related to the mass and spin of the post-merger black hole \cite{QNM}. Improved frequency sensitivity in the kilohertz range allows for better detection of lower mass black hole quasi-normal modes. Thus, in both cases, improved high frequency sensitivity is important.

The sensitivity of LIGO-type laser interferometers degrades at high frequency due to quantum shot noise and the bandwidth of the optical cavities. The signal recycling mirror (SRM) used for resonant enhancement of the of the signal can strongly modify the response of the detector, allowing it to be tuned to narrow or broadband enhancement \cite{BuonnanoScaling, BuonnanoNoise}.

Two methods have been proposed for improving the sensitivity in the 1-4 kHz range. The first is to use high optical power with optical squeezing \cite{KimbleSqueezing}. Currently, phase noise squeezing has been demonstrated in GEO \cite{GEOSqueezing} and Advanced LIGO \cite{Aasi2013}. Frequency dependent squeezing of quantum noise has been demonstrated in a tabletop setting \cite{OelkerFD}, which will allow improvement at both high and low frequencies.  Vahlbruch, \textit{et al.} demonstrated detection of a 15 dB squeezed vacuum state using a photodetector with 0.995 quantum efficiency \cite{SchnabelSqueezer}. The primary limitation on sensitivity improvement from squeezing is optical losses in the interferometer and injection optics, which degrade the level of squeezing. 

The second proposed approach for enhancing high frequency sensitivity is through white light signal recycling (WLSR). This entails placing a negative dispersion filter within the signal recycling cavity so that the GW signal sideband frequency dependent phase delay is cancelled, thereby allowing a broad range of frequencies to be simultaneously resonant. The basic concept was proposed in by Wicht, \textit{et al.} and demonstrated using atomic media in 1999 and 2007 \cite{AkulshinWhiteLight, WangWhiteLight, PatiWhiteLight}. However, Ma \textit{et al.} showed that atomic media introduced unacceptable noise into GW interferometers \cite{YiqiuWhiteLight}. Miao et. al showed that unstable optomechanical filters can also be used to create negative dispersion, and proposed that the intrinsic instabilities can be controlled by linear feedback \cite{HaixingFilters}. Qin, \textit{et al.} demonstrated the principle of linear negative dispersion using a blue detuned optomechanical filter \cite{JiayiND}.

The experimental challenge of creating a low noise WLSR interferometer is the stringent requirement on thermal noise which scales as \(T\cdot Q_m^{-1}\), where \(T\) is the environmental temperature and \(Q_m\) is the mechanical quality factor (Q-factor). Quality factors $\gtrsim$ 10$^{10}$ are required, which is several orders of magnitude higher than most low loss materials. However, Q-factor enhancement is possible using optical dilution, in which radiation pressure forces are used to replace the mechanical restoring force of the resonator, diluting the effect of mechanical dissipation. This can in principle allow the Q-factor to be enhanced sufficiently to meet the thermal noise requirements, assuming a cryogenic operating temperature \cite{YiqiuDilution, PageCatflap}. 

Optical dilution has been demonstrated using a gram-scale resonator \cite{CorbittDilution}, and with optically trapped microresonators \cite{CorbittTrap, NiTrap}. However, optical dilution is limited by acceleration loss at high frequencies \cite{ScienceChina}. A particular implementation known as the \textit{cat-flap resonator} was discussed in reference \cite{PageCatflap}. It was shown that Q-factors as high as $10^{14}$ could in principle be achievable using mm-scale resonators suspended by atom-scale membranes such as graphene sheets or nanowires.

White light signal recycling is compatible with optical squeezing, as long as the optical losses introduced by the filter cavity are kept below that of the dominant loss source of the injection and interferometer. Thus both methods can be combined to achieve optimal performance. In this paper, we consider currently realizable technologies for implementing WLSR based on low loss AlGaAs/GaAs resonators. Exploration of the parameter space reveals a small regime in which the optical dilution of the level required can be achieved in practical optical devices. In section \ref{sectionNDF}, we review the theoretical background of the WLSR, from which we identify the frequency and Q-factor requirements for creating a suitable negative dispersion filter. In section \ref{sectionCFR} we analyze the optical dilution scheme for cat-flap resonators, and use finite element analysis to refine the cat-flap resonator dimensions. All of the results are combined in section \ref{sectionDIS} in which we estimate how the sensitivity of an Advanced LIGO type GW detector could be modified by replacing the signal recycling cavity with a WLSR cavity.

\begin{figure}\begin{center}
\includegraphics[width=\textwidth]{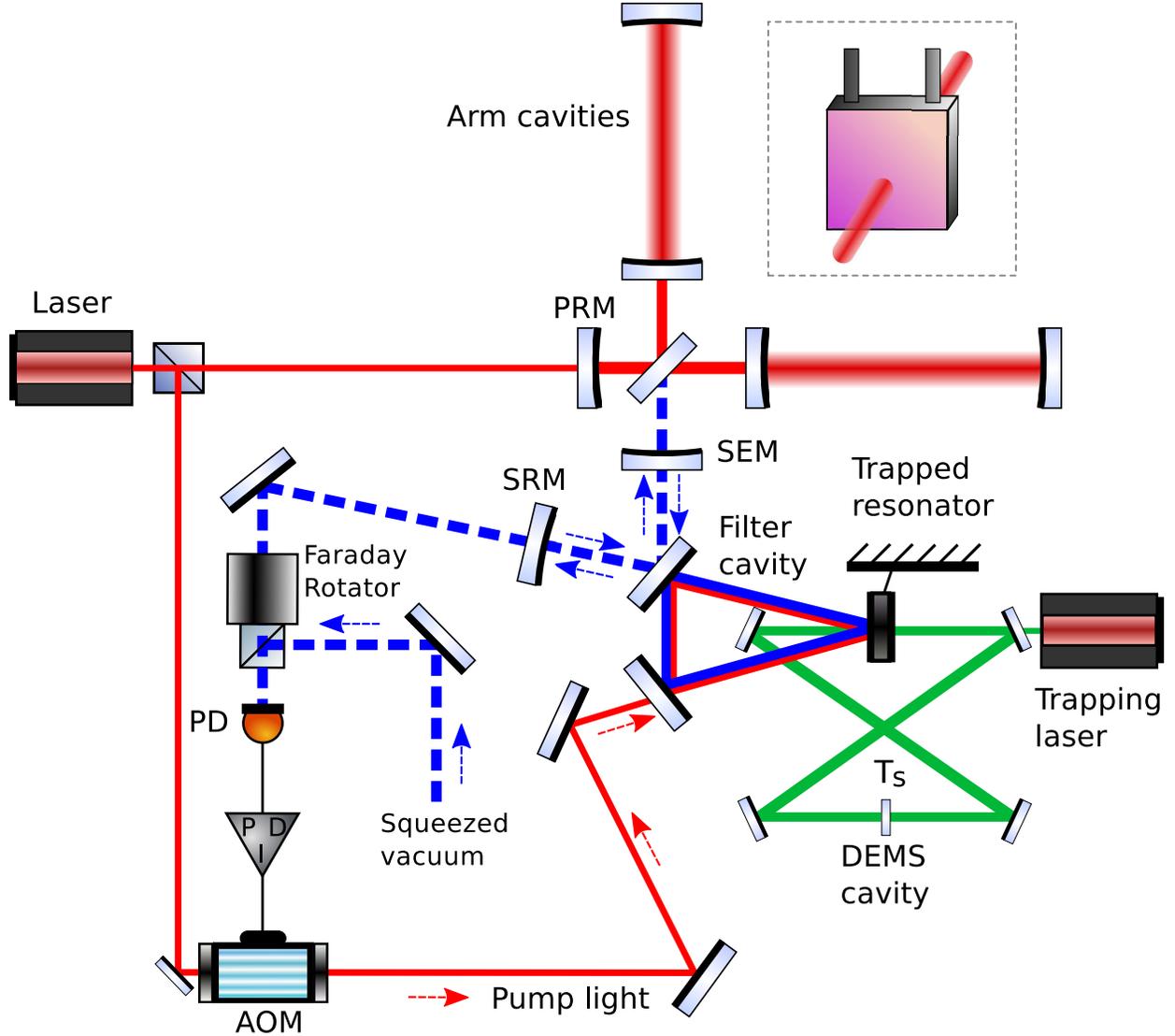}
\end{center}
\caption{Proposed setup for a white light signal recycling (WLSR) interferometer. The optically trapped cat-flap resonator acts as the mechanical component for both the bowtie Double End Mirror Sloshing (DEMS) cavity and the triangular negative dispersion filter cavity. Feedback forces are applied by intensity modulation of the pump light using an acousto-optic modulator (AOM) driven by a PID control loop that provides intensity modulation of the blue detuned pump light that drives the filter cavity. PRM, SRM and SEM denote power recycling mirror, signal recycling mirror and signal extraction mirror. $T_s$ denotes the DEMS cavity sloshing mirror with transmissivity $\sim 300$ ppm.  Squeezed vacuum is injected via the output Faraday isolator.  \textbf{Inset:} A ``cat-flap'' mechanical resonator. The entire structure is a high reflectivity AlGaAs/GaAs coating, suspended by thin ribbons created by cutting or etching of the coating. \label{WLC}}
\end{figure}

\section{Negative dispersion filter}\label{sectionNDF}

We consider a dual recycling interferometer in which the signal recycling mirror is replaced by a signal extraction mirror and a negative dispersion filter cavity, which includes an optically diluted mechanical resonator achieved using a double end mirror sloshing cavity shown in figure \ref{WLC}. Technical details will be discussed later in section \ref{sectionCFR}.

First, we shall review the concept of the negative dispersion optomechanical filter. Gravitational waves of frequency $\Omega$ interact with an interferometer to generate sidebands to the carrier light. These sidebands have a positive phase delay, which can be cancelled by passage through a filter cavity that applies a frequency dependent negative phase, known as negative dispersion. Miao, \textit{et al.} showed that negative dispersion can be generated by optomechanical interactions in the presence of a strong blue-detuned pumping field at frequency \(\omega_c + \omega_{m}\), where $\omega_c$ is the filter cavity resonant frequency and \(\omega_{m}\) is the mechanical resonant frequency \cite{HaixingFilters}. The Hamiltonian which describes the system is given by $\hat H=\hbar (\omega_c+ g_{0}\hat x)\hat a^{\dag}\hat a+\hat H_{m}+\hat H_{\gamma}$. Here, $\hat H_{m}=\hat p^2/2m+m\omega_{m}^2\hat x^2/2$ is Hamiltonian of the mechanical oscillator. $\hat H_{\gamma}$ describes the interaction between the intra-cavity field $\hat a$ and external optical fields. $g_{0}$ is the linear optomechanical coupling strength. In the rotating frame at frequency $\omega_c$, we have:

\begin{subequations}
\begin{align}
&\hat{x}(\omega_m - \Omega)=\chi_m\{-\hbar g_{0}\bar {a}[\hat{a}(\Omega)+\hat{a}^{\dag}(2\omega_m-\Omega)]+\xi_{\rm th}\},\\
&\hat{a}(\Omega)=\chi_c[-i g_{0}\bar {a}\hat{x}+\sqrt{2\gamma_f}\hat{a}_{\rm in}(\Omega)],
\end{align}
\end{subequations}
\noindent
where $\chi_m = -m[(\omega_m - \Omega)^2-\omega_m^2+i\gamma_m (\omega_m-\Omega)]^{-1}$ is the mechanical susceptibility, $\chi_c=[-i\Omega+\gamma_f]^{-1}$ is the optical susceptibility, and $\xi_{\rm th}$ represents the thermal fluctuations. \(\gamma_{f}\) is the unmodified filter cavity bandwidth and $\gamma_m$ is the initial mechanical damping.

In order to keep the upper sideband $\hat{a}^{\dag}(2\omega_m-\Omega)$ out of resonance, we require the resolved sideband condition to be met such that $\omega_m\gg\gamma_f\gg\Omega $. This requirement on $\omega_m$ will be discussed further in section \ref{sectionCFR}. Here we note that to maintain effective negative dispersion we need $\omega_m/(2\pi) >$ 100 kHz. Under the resolved sideband condition, the mechanical and optical susceptibility can be approximated as $\chi_m=[2m\omega_m(\Omega+i\gamma_m)]^{-1}$ and $\chi_c\sim\gamma_f^{-1}$ respectively. Additionally, the mechanical susceptibility approximation requires the condition that the detuning of the pump field is close to $\omega_m$. According to the input-output relation $\hat{a}_{\rm out}=-\hat{a}_{\rm in}+\sqrt{2\gamma_f} \hat a$, the input-output response of the negative dispersion filter is given by:

\begin{equation}\label{ndfio}
\hat{a}_{\rm out}(\Omega) = \frac{\Omega + i \gamma_{\rm opt}}{\Omega - i \gamma_{\rm opt}}\hat{a}_{\rm in}(\Omega) + \hat{n}_{\rm th}(\Omega),
\end{equation}
\noindent
where $\gamma_{\rm opt}=\hbar g_0^2\bar a^2/2m\omega_m\gamma_f$ is the optomechanical damping bandwidth \cite{HaixingFilters}.

The first term gives the negative optical response, with a filter phase \(\phi_f \sim -2i\Omega/\gamma_{\rm opt}\). The negative dispersion is approximately linear when \(\Omega\ll\gamma_{\rm opt}\). Here, the optomechanical damping \(\gamma_{\rm opt}\) can be tuned by changing the pumping field power. The filter phase $\phi_f$ must be matched to the arm cavity delay, which makes the optimal $\gamma_{\rm opt}/(2\pi)$ approximately 12 kHz. 

The second term \(\hat{n}_{\rm th}\) represents the thermal fluctuation of the mechanical oscillator. The requirement for the thermal noise to be lower than the quantum noise in the interferometer is \cite{HaixingFilters}:

\begin{equation}\label{TQSR}
8 k_{\rm B} T\cdot Q_m^{-1} \lesssim \hbar \gamma_{\rm sr, eff},
\end{equation}
\noindent
where \(k_{\rm B}\) is the Boltzmann constant and $\gamma_{\rm sr, eff} $ is given by the normal expression for the bandwidth of a signal recycled interferometer \cite{StefanReview}:
\begin{equation}
\gamma_{\rm sr, eff} = \frac{c T_{\rm sr}}{4 L_{\rm arm}},
\end{equation}
\noindent
where \(T_{\rm sr}\) is the signal recycling mirror transmissivity and $L_{\rm arm}$ is the interferometer arm length.

The general configuration of a WLSR interferometer is dictated by the need for coupling the optomechanical filter to the interferometer while simultaneously implementing an optical dilution scheme to allow equation \ref{TQSR} to be satisfied. The optical dilution can be realized using a Double End Mirror Sloshing (DEMS) cavity \cite{PageCatflap} that is further discussed in section \ref{sectionCFR}. The DEMS cavity reduces quantum radiation pressure noise and produces a stable optical trap \cite{YiqiuDilution} in which the mechanical resonance $\omega_m$ is shifted to the optical spring frequency $\omega_{\rm opt}$. The configuration in figure \ref{WLC} combines the negative dispersion filter with optical dilution. The WLSR filter now plays the role of the signal recycling mirror in a conventional gravitational wave detector. Note that squeezed vacuum can be injected at the output Faraday rotator in the same manner that has been demonstrated in conventional interferometers \cite{Aasi2013,GEOSqueezing}. Squeezed vacuum of 15 dB has been measured with 0.995 quantum efficiency photodetectors \cite{SchnabelSqueezer}, and combined with improvements in the interferometer optical loss, it is assumed that a resultant frequency dependent squeezing magnitude of 10 dB is possible.

\begin{figure}\begin{center}
\includegraphics[width=\textwidth]{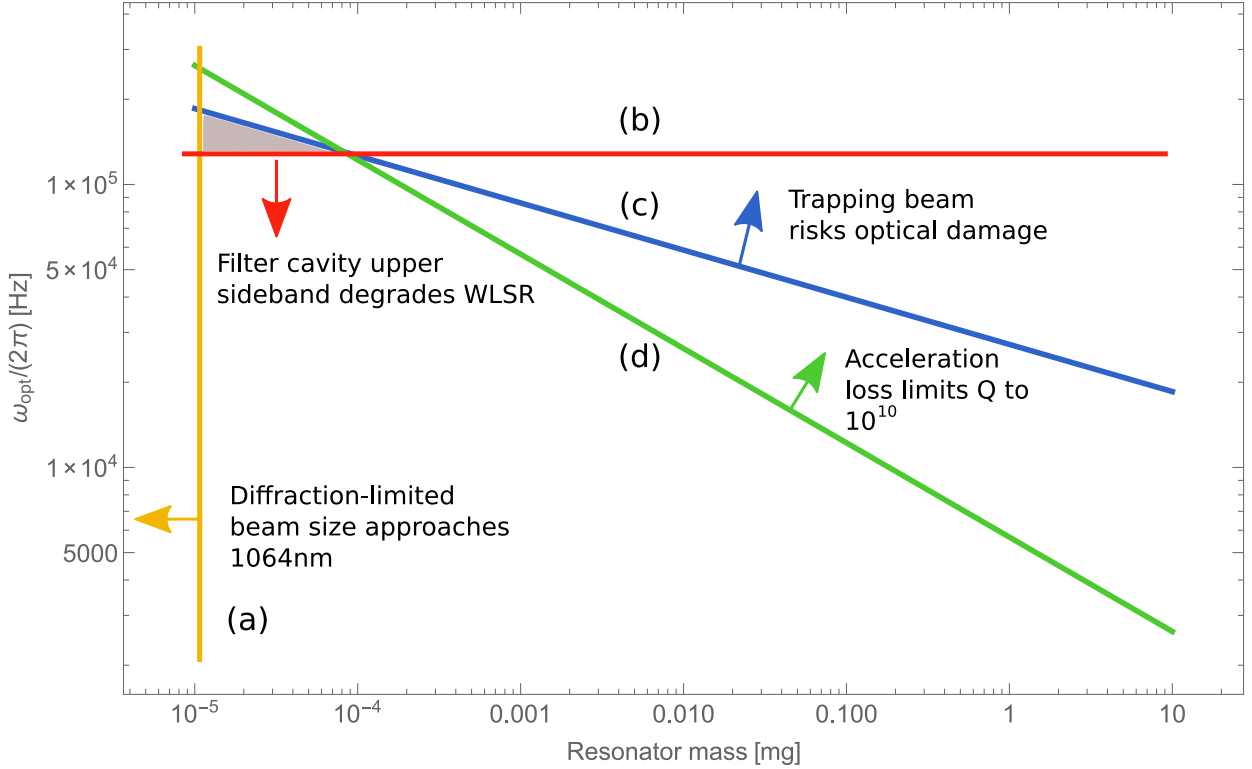}
\end{center}
\caption{Approximation of the parameter space for constructing a white light cavity with a negative dispersion filter, DEMS cavity, AlGaAs/GaAs cat-flap resonator and Advanced LIGO design power and test mass parameters. \textbf{(a):} The lower limit of the mirror size is determined by diffraction loss of 1ppm. If the mirror is too small, the required beam size approaches the wavelength. \textbf{(b):} The lower bound on $\omega_{\rm opt} \sim \omega_m$ is set by the resolved sideband condition $\omega_m \gg \gamma_f$. \textbf{(c):} Optical spring frequency versus mass in a DEMS cavity, at the power which reaches the lower bound  AlGaAs/GaAs coating damage of 4 MW/cm$^2$ \cite{UFLDamage}. \textbf{(d):}\label{pSpace} Optical spring frequency at which the Q-factor cannot exceed $10^{10}$ due to acceleration loss, which is calculated with respect to the first flexural mode of a bending plate that is free at all edges. This is discussed in detail in section \ref{sectionCFR}.}
\end{figure}

It is instructive to consider the parameter space for the design of a WLSR system. We calculate the mirror mass against optical spring frequency for possible WLSR setups, the results of which are shown in figure \ref{pSpace}. The mirror geometry is kept as a square prism with thickness one quarter of the side length, so that the mirror size defines the mass. Diffraction loss must be kept to 1 ppm, which requires using a beam radius one-fifth of the mirror's side length. This requirement sets the lower limit of mass, since the beam size must be larger than the assumed 1064 nm wavelength. The minimum optical spring frequency is set by the resolved sideband condition of the negative dispersion filter, which is shown as a horizontal line at 120 kHz in figure \ref{pSpace}. The sloped lines in figure \ref{pSpace} show the acceptable levels of optical intensity and acceleration loss, both of which are further discussed in section \ref{sectionCFR}. The acceleration loss limit is dictated by equation \ref{TQSR}. Assuming operation at an environmental temperature of 4 K, the required Q-factor is $Q_m \gtrsim$ 10$^{10}$. The parameter curve for the corresponding acceleration loss was calculated using the plate bending mode of the mirror. The maximum optical intensity is set by the lower bound of AlGaAs damage threshold \cite{UFLDamage}. However, it is important to note that this value represents the highest intensity applied to AlGaAs mirrors without observing optical damage, and thus the actual intensity limit is higher, but currently unknown.  The shaded region in figure \ref{pSpace} meets the requirement that a practical configuration needs to be able to provide an 8-fold strain sensitivity improvement at 2 kHz, without introducing unacceptable noise at 100 Hz. Thus we see that resonators of mass 10-100 ng are required for practical WLSR. Somewhat better performance could be achieved if the damage threshold of AlGaAs is found to be higher.

\section{Cat-flap resonator design requirements}\label{sectionCFR}

The cat-flap resonator is mounted in a double end mirror sloshing (DEMS) cavity, where it acts as the end mirror of two cavities coupled by a partially transmitting sloshing mirror, shown in the inset of figure \ref{WLC}. The DEMS cavity mimics a membrane-in-the-middle cavity, but allows the use of mechanical resonators of arbitrary thickness \cite{PageCatflap}. Our estimates of the resonator losses relate to details of device fabrication which we summarize below. 

The cat-flap resonators using focused ion-beam fabrication of AlGaAs/GaAs coatings. The coatings are bonded to a silicon wafer on which windows of varying sizes have been patterned. Focused ion beam machining allows fabrication of single monolithic resonators, to be discussed in future papers. Each resonator consists of a rectangular mirror suspended by thin ribbons of coating material. We used a combination of analytic approximations and finite element analysis to model the losses. First, we shall review the key loss terms, and then describe the finite element analysis used to model the stress and strain distributions of the cat-flap as a function of frequency. This result is used as a measure of the thermal noise coupling of the optomechanical unstable filter, which enables prediction of the strain sensitivity curve of an Advanced LIGO detector in which WLSR replaces conventional signal recycling.

 \subsection{Optical dilution and acceleration loss}
 
Two main factors contribute to the losses in optically diluted resonators. The first factor describes the loss reduction due to optical dilution, and the second factor describes the losses associated with acceleration through an optical spring. Optical dilution suppresses mechanical dissipative losses relative to non-dissipative optical losses, according to the ratio of the contribution of the optical and mechanical spring constants. In \cite{PageCatflap} it was shown that:
\begin{equation}\label{osDilution}
Q_{\rm opt}^{-1} \sim \frac{\omega_{0}}{\omega_{\rm opt}}Q_0^{-1}.
\end{equation}

Here $Q_{\rm opt}$ is the optically diluted Q-factor and $Q_0$ is the mechanical Q-factor in the absence of radiation pressure. The zero gravity resonance frequency $\omega_0$ is determined by the mechanical spring stiffness and hence the noise coupling to the thermal reservoir. Although the device is normally suspended like a pendulum, the gravitational spring constant does not introduce dissipation,  and is negligible compared to the optical spring constant. 

In general, in order to suppress the mechanical dissipation, it is clearly advantageous to use the softest possible suspension. In \cite{PageCatflap} it was shown that suspensions with atomic dimensions  could achieve $\omega_0/(2\pi) =$ 1 Hz. However, we will see in section \ref{subsectionMOD} that such soft suspensions are not necessary for the negative dispersion filter, because the $\omega_{\rm opt}$ requirement places the device in the regime where acceleration loss dominates. 

Acceleration loss arises due to deformation of the mirror when accelerated by optical forces. In reference \cite{PageCatflap} a 1 dimensional approximation was used to estimate acceleration loss of a resonator with a single mechanical mode $\omega_{\rm int}$. It was shown that the optically diluted Q-factor was bounded by an acceleration loss limited $Q_{\rm acc}$:
\begin{equation}\label{osA}
Q_{\rm acc}^{-1} \sim \frac{\omega_{\rm opt}}{\omega_{\rm int}}Q_{\rm int}^{-1},
\end{equation}
where $Q_{\rm int}$ is the Q-factor of the mirror mechanical mode. This means that the resonator should have internal mode frequencies that are much larger than the desired optically diluted resonant frequency. This in turn leads to a requirement for resonators that are small, with mirror thickness comparable to the other dimensions. In section \ref{subsectionMOD} we use finite element analysis to make a detailed estimate of this term, which also takes into account contributions from multiple acoustic modes of the mirror and suspension.

\subsection{Gas damping, thermoelastic loss and absorption heating}\label{subsectionTHM}

Residual gas pressure limits the obtainable Q-factor to a value given by  \cite{PressureQ}:
\begin{equation}\label{gasDamp}
Q_{\rm gas} = \frac{\omega_{\rm opt} \eta}{4 P_{\rm air}}\sqrt{\frac{\pi R_{\rm gas} T}{2 M_{\rm air}}}
\end{equation}
where $\eta$ is the mass per unit area of the mirror, normal to the direction of vibration, $P_{\rm air}$ is the residual pressure, $R_{\rm gas}$ is the universal gas constant and $M_{\rm air}$ is the molar mass of air. It will be shown in section \ref{subsectionMOD} that a resonator thickness of 10 $\mu$m is required. Equation \ref{TQSR} dictates that we must achieve a Q-factor of the order of $10^{10}$. Using the cat-flap resonator parameters, equation \ref{gasDamp} dictates that a pressure of $10^{-8}$ Torr is sufficient to keep the gas damping limited Q-factor above $10^{11}$.

Thermoelastic loss is a critical factor for the design of the resonator and determination of the operating temperature. Thermoelastic effects scale with the thermal expansion coefficient and can be suppressed in materials such as GaAs in which the thermal expansion coefficient has zero crossings at 55 K \cite{AdachiGa} and 13 K \cite{ColeComms, IoffeInstitute}. In addition, the thermal expansion coefficient of AlGaAs/GaAs coatings is less than 10$^{-8}$ at temperatures below 13 K \cite{ColeComms}. Thermoelastic loss is approximated by the following \cite{ZenerThermo, ZenerThermo2, LifshitzThermo}:

\begin{equation}\label{thermoelasticRevised}
\phi_{\rm TE} = \frac{Y \alpha^2 T}{C_p}\frac{\omega_{\rm opt}\tau}{1+ \omega_{\rm opt}^2\tau^2}
\end{equation}

\noindent
where \(Y\) is Young's Modulus, \(\alpha\) is the coefficient of thermal expansion, \(C_p\) is the specific heat at constant pressure and \(\tau\equiv(t_{\rm flex}^2 C_p) / (\pi^2 \kappa)\), where \(\kappa\) is the thermal conductivity and $t_{\rm flex}$ is the thickness of a flexural member. It is possible to reduce the thermoelastic loss in both the suspension and mirror to a negligible amount by cooling to the temperature where there is zero thermal expansion. For AlGaAs/GaAs coatings, $\alpha<10^{-8}$ for temperatures less than 13 K \cite{IoffeInstitute, ColeComms}. At 4 K, \(\kappa \sim \) 150 W/(m.K) and \(C_p \sim\) 2 kJ/(m\(^3\).K), giving a thermoelastic loss angle contribution of $\phi_{\rm TE} < 10^{-12}$ in the mirror and $\phi_{\rm TE} < 10^{-16}$ in the suspension.

\begin{figure}\begin{center}
\includegraphics[width=\textwidth]{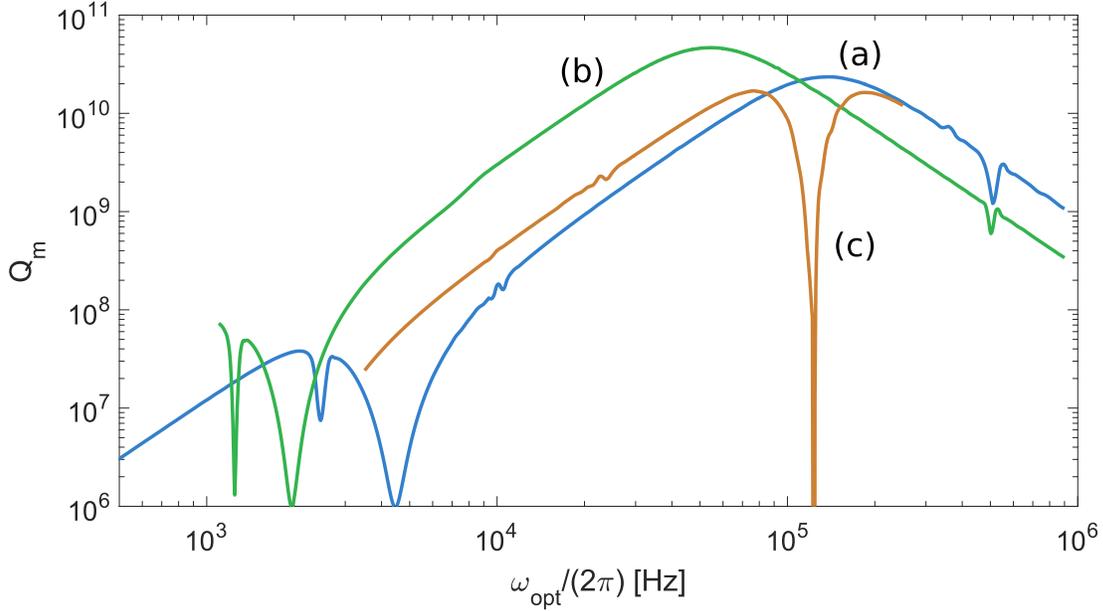}
\end{center}
\caption{Three examples of optical dilution enhanced Q-factor $Q_m$ versus optical spring frequency, calculated using finite element analysis discussed in section \ref{subsectionMOD}. \textbf{(a)}: The resonator mirror is a 30 $\mu$m square prism with 10 $\mu$m thickness. The suspension is comprised of two ribbons, each 15 $\mu$m length, 6 $\mu$m width, 25 nm thickness. The features at 2.5 kHz and 4.5 kHz are torsion and rocking modes. The feature  at 500 kHz is the first transverse suspension mode. \textbf{(b)}: Resonator size increased to 50 $\mu$m square prism with 10 $\mu$m thickness, same suspension dimensions as (a). The value of $\omega_0$ is lower than that of (a), thus increasing the dilution factor, but the increased acceleration loss degrades $Q_m$ at $\omega_{\rm opt}/(2\pi) =$ 170 kHz. The torsion and rocking modes are located at 1.1 kHz and 2 kHz. The first transverse mode is near 500 kHz. Comparison with theoretical predictions indicate that these transverse modes are bending modes of a fixed/fixed beam, rather than tensile string modes. Thus, they do not change significantly with the mass of the mirror when the width of the suspension is comparable to the size of the mirror. \textbf{(c)}: mirror is the same size as in (a), but the suspension is lengthened to 30 $\mu$m. This lowers $\omega_0$ but also causes a problematic transverse suspension mode near $\omega_{\rm opt}/(2\pi) =$ 170 kHz. Longer suspension also does not improve the modelled $Q_m$ in the acceleration loss limited regime. Note that small features at 10 kHz, 20 kHz and 350 kHz in all three curves are interpolation errors where modelling frequency bands are joined. \label{AccelLossQ}}
\end{figure}

The thermal heating of the cat-flap resonator is determined by the total power in the cavity. The circulating power in the DEMS cavity required to achieve a certain optical spring frequency \(\omega_{\rm opt}\) is given by \cite{YiqiuDilution, PageCatflap}:

\begin{equation}\label{powerTrap}
P_{\rm trap} = \frac{\omega_d \omega_s \omega_{\rm opt}^2 m \gamma_d}{g_0^2 T_f},
\end{equation}

\noindent
where \(\omega_s = c\sqrt{T_s}/2L\) is the sloshing frequency, \(T_s\) is the sloshing mirror transmissivity, $T_f$ is the input mirror transmissivity and $\omega_d$ and $\gamma_d$ are the optical frequency and bandwidth of the DEMS cavity. To reduce the circulating power, we wish to minimize the mirror mass and $T_s$. As shown in figure \ref{pSpace}, the lower limit of the resonator size is set by diffraction loss requirements. The lower limit of $T_s$ is set by the fact that the DEMS cavity radiation pressure noise cancellation leaves residual radiation pressure noise that is inversely proportional to $T_s$ \cite{YiqiuDilution}. Setting the requirement that the residual noise force be below the thermal noise force of the cat-flap, we find that $T_s$ = 300 ppm. 

The equilibrium temperature of the mirror can be estimated by treating conduction through the suspension as the limiting process of heat transfer. We use a 1-D conduction model to calculate the temperature drop across the ribbons, which has also been shown in the context of cryogenic optically diluted resonators by Ma \cite{YiqiuThesisHeat}. The power conducting through the ribbons is given by:
\begin{equation}\label{HC1D}
P_{\rm cond} = - \sigma \kappa(T) \frac{dT}{dz}
\end{equation}
 where $\sigma$ is the cross sectional area of the ribbons, $\kappa(T)$ is the temperature dependent thermal conductivity and $z$ is the coordinate in the direction of conduction. At cryogenic temperature, $\kappa$(T) can be approximated by $\kappa_0\times$T$^{n}$. For AlGaAs/GaAs mirrors below 10 K, $\kappa_0$ = 4.81 W/(m.K$^{n}$) and n = 2.3 \cite{ColeComms}. Equating $P_{\rm cond}$ with $P_{\rm abs}$, the power absorbed by the mirror, and integrating equation \ref{HC1D} gives:
\begin{equation}\label{eqT}
P_{\rm abs} = \kappa_0\times\frac{2 \sigma}{(n+1) l_{\rm sus}}(T_m^{n+1}-T^{n+1}),
\end{equation}
 where $T_m$ is the equilibrium temperature of the mirror in the limit of conduction heat transfer and $l_{\rm sus}$ is the length of the suspension ribbons. It is assumed that the supporting structure can be maintained at an environmental temperature of T = 4 K. Using absorption of AlGaAs approximately 0.7ppm \cite{ColeCrystalline2}, and ribbon dimensions that will be discussed in section \ref{subsectionMOD}, it would take 5 W of circulating power to cause a temperature drop of 1 K across the ribbons. This exceeds the 2 W circulating power required for a 170 kHz optical spring with a 40 ng resonator. In this case, the beam intensity is up to 2.5 MW/cm$^2$, where the laser beam radius is limited by the side length of the resonator in order to reduce diffraction loss to 1 ppm. AlGaAs/GaAs crystalline coatings have been shown to withstand 4 MW/cm$^2$ intensity at 1064nm without damage \cite{UFLDamage}.

\subsection{Finite Element Analysis}\label{subsectionMOD}

Finite element analysis in COMSOL 5.3 is used to estimate the optically diluted Q-factor with the presence of suspension losses combined with other losses discussed above. A geometry independent AlGaAs/GaAs loss angle of 10$^{-6}$ is assumed. The ribbons have an initial tensile stress due to the hanging mirror. A fixed constraint is applied at the top of the ribbons. Deformation is assumed to be linear and elastic, calculated with the average Young’s Modulus, density and Poisson ratio of AlGaAs/GaAs coatings. The optical spring is applied at the centre of percussion to minimize suspension loss \cite{BraginskyPercussion, PAGE2017}. This spring is set as a restoring force -k$_{\rm opt}\cdot$\textbf{v}, where \textbf{v} is the displacement in the direction normal to vibration. The structure is meshed much finer in the ribbons than in the mirror since the elements must not be excessively thin. During the solution process, the spring constant k$_{\rm opt}$ is swept through values of 10$^{-8}$ to 10$^{-1}$ N/m. An eigenvalue transform point must be set, otherwise the solution becomes imprecise at high k$_{\rm opt}$. However, this cuts off all eigenfrequencies below the selected value. Thus, the solution is obtained in frequency bins that approximately span an order of magnitude in the frequency domain. This leads to small interpolation errors where these sections have been joined. The solver produces a list of eigenfrequencies with an associated Q-factor. The result of this modelling is shown in figure \ref{AccelLossQ}. The mode associated with $\omega_0$ is the fundamental bending mode of a fixed/free cantilever with an end mass. This mode increases in resonant frequency and Q-factor as the optical spring constant is increased.

Acceleration loss from coupling to the mirror’s internal modes becomes significant at optical spring frequencies approaching 100 kHz. This acceleration loss is reduced by having a small mirror. The resonator in figure \ref{AccelLossQ}(a) has the highest Q-factor at $\omega_{\rm opt}/(2\pi) =$ 170 kHz compared to the other modelled cases. As discussed above, the lower limit of the mirror size is set by diffraction loss and beam size considerations. The thickness must also be suitably high to allow the mirror to be highly reflective.

Damping in the ribbons reduces the $Q_m$ across the entire frequency range, which sets the requirement of having low ribbon thickness. The yield stress condition is not problematic for these microribbons as the resonators in figure \ref{AccelLossQ} have a ribbon tensile stress of $\sim$ 2 kPa. It can also be seen in figure \ref{AccelLossQ}(a) that suspension modes do not interfere with the operation of the optical spring at 170 kHz. 

\subsection{Strain sensitivity prediction}\label{subsectionSEN}

The results obtained above can now be used to estimate the performance of an Advanced LIGO type interferometer with a WLSR cavity that uses the resonator described in figure \ref{AccelLossQ}(a) above. The Q-factor at 170 kHz is extracted from the finite element results using interpolation, and the result is entered as a parameter in the input output relation of the filter cavity shown in equation \ref{ndfio}. The calculation considers quantum noise of the interferometer and filter using the two photon formalism established by Caves and Schumaker \cite{CavesTwoPhoton1, CavesTwoPhoton2}, which was adapted to the context of gravitational wave detectors by Corbitt, \textit{et al.} using a transfer matrix method \cite{CorbittTwoPhoton}. We also consider thermal noise from the cat-flap resonator, quantified by $T\cdot Q^{-1}_{\rm m}$. Similar calculations carried out regarding quantum and thermal noise of WLSR unstable filters have been discussed in references \cite{HaixingFilters, JiayiThesis}. We assume that the round trip optical loss of the filter cavity is less than 10 ppm. This requires extremely low absorption, which is possible using crystalline AlGaAs mirrors, and  1 ppm or less of diffraction loss on the cat-flap resonator. 

\begin{figure}\begin{center}
\includegraphics[width=\textwidth]{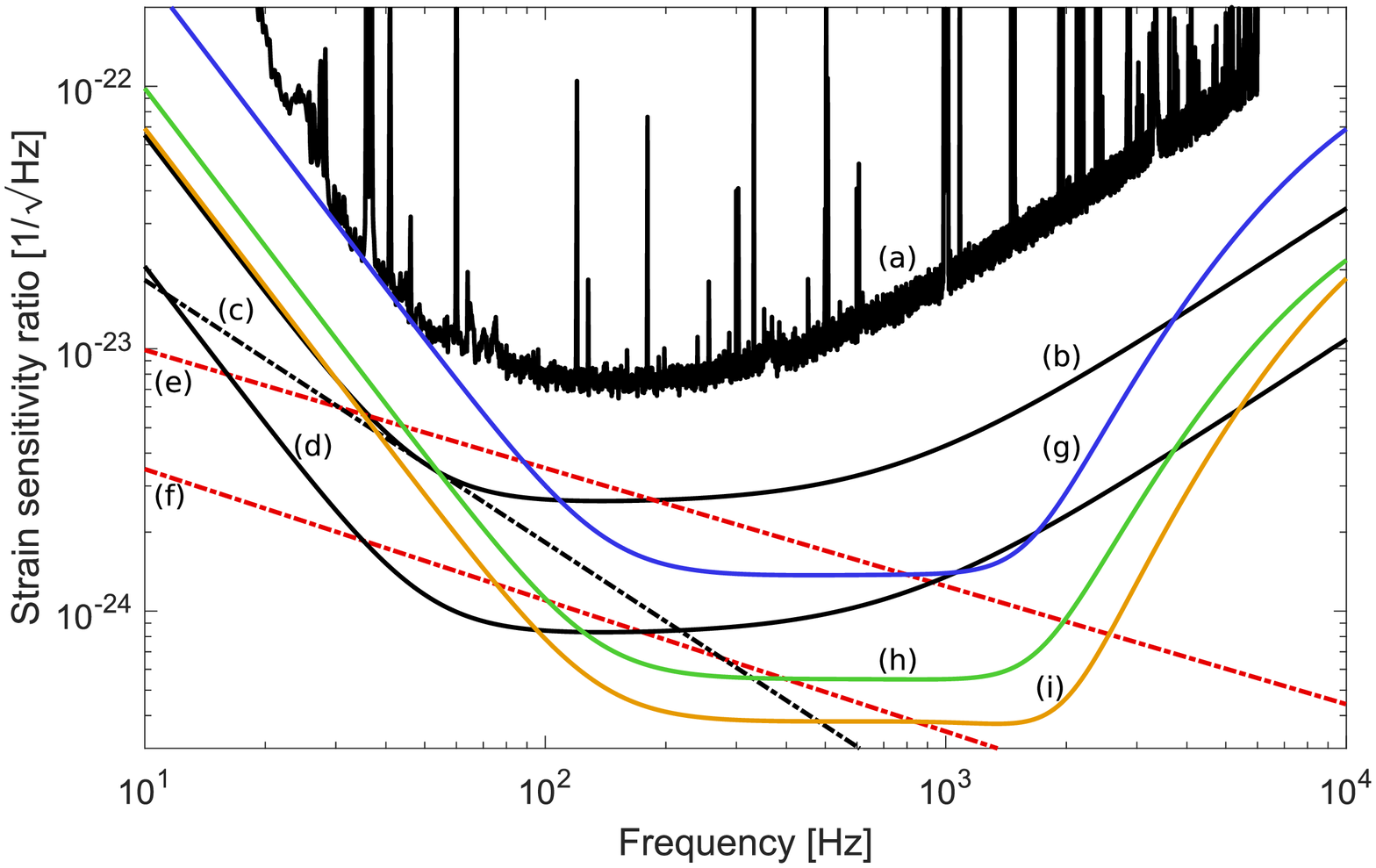}
\end{center}
\caption{Strain sensitivity curves. All calculations use 40 kg test masses and 4 km arms. All curves except (a) use 800 kW circulating power. \textbf{(a)} Hanford O1 with a circulating power of 100 kW \cite{HanfordSens}. \textbf{(b)} Advanced LIGO High Power quantum noise with broadband tuning. \textbf{(c)} Advanced LIGO standard quantum limit. \textbf{(d)} Advanced LIGO high power quantum noise with 10 dB resultant frequency dependent squeezing. \textbf{(e)} Room temperature Brownian noise of Advanced LIGO fused silica test masses with silica/tantala coatings \cite{SlawekCBTN}. \textbf{(f)} Room temperature Brownian noise of Advanced LIGO test masses with crystalline AlGaAs/GaAs coatings \cite{ColeCoherent}. \textbf{(g)} WLSR enhancement with 170 kHz optical spring, $\gamma_f/(2\pi) =$ 2 kHz and \(T\cdot Q_m^{-1} = 2 \times 10^{-10}\), using the resonator described in this paper. \textbf{(h)} WLSR from curve (g) combined with 10 dB resultant frequency dependent squeezing. \textbf{(i)} Idealized WLSR with 10 dB frequency dependent squeezing, using a resonator operating at 500 kHz optical spring frequency, \(T\cdot Q_m^{-1} = 4 \times 10^{-11}\) and sub-ppm filter cavity optical loss.\label{LIGOSens}}
\end{figure}

White light signal recycling has been calculated and compared to various strain sensitivity curves for Advanced LIGO, the results of which are shown in figure \ref{LIGOSens}. For the rest of this subsection, letters in parenthesis refer to figure \ref{LIGOSens} only. Curves (a)-(f) are shown to give context before introducing the white light signal recycling. The measured sensitivity of the Hanford site during its O1 run is shown by (a) \cite{HanfordSens}. The Advanced LIGO quantum noise sensitivity, tuned for broadband detection, is shown by (b), which touches the standard quantum limit (c) at approximately 50 Hz. A 10 dB reduction of the quantum noise across the entire frequency band is shown by curve (d), which represents frequency dependent squeezing. Gras and Evans measured the room temperature Brownian noise spectrum of Advanced LIGO silica/tantala coatings, and used this measurement to calculate the Brownian noise of the Advanced LIGO test masses shown by (e) \cite{SlawekCBTN}. The current coating Brownian noise of Advanced LIGO is above the quantum noise of (b) for frequencies up to 200 Hz. Chalermsongsak, \textit{et al.} designed a specialized AlGaAs/GaAs crystalline coating to minimize the coating Brownian and thermo-optic noise, and calculated the respective thermal noises of Advanced LIGO fused silica test masses with the optimized coatings \cite{ColeCoherent}. The Brownian noise of these improved test masses is shown by curve (f), which represents more than a factor of 2 improvement versus (e). The thermo-optic noise of the improved test masses has a value of 3 $\times$ 10$^{-25}$ Hz$^{-1/2}$ at 100 Hz and remains less than this value between 100 Hz and 10 kHz \cite{ColeCoherent}. 

Curves (g)-(i) incorporate white light signal recycling. The optically diluted cat-flap resonator that follows the design specifications outlined in section \ref{subsectionMOD} produces WLSR characterized by (g). As discussed in section \ref{sectionNDF} and shown in figure \ref{WLC}, WLSR can be combined with frequency dependent squeezing. Applying 10 dB reduction of quantum noise to (g) gives the enhancement shown by (h). It is seen that this WLSR configuration produces an 8-fold enhancement at 2 kHz detection frequency relative to the Advanced LIGO quantum noise. By comparison with (e) and (f), significant improvement in the Advanced LIGO coating Brownian noise is required to reach this sensitivity. Curve (i) shows the improvement possible if the $T\cdot Q^{-1}_{\rm m}$ could be improved by a factor of 5 compared to our WLSR design, while also operating at 500 kHz optical spring frequency, and implementing further improvements to the test mass coating thermal noise. Lowering the thermal noise and operating further in the resolved sideband regime reduces noise floor and increases the bandwidth of WLSR.

\section{Discussion}\label{sectionDIS}

The restrictions on the design described throughout this paper lead us to select dimensions of the cat-flap listed for curve \textbf{(a)} of figure \ref{AccelLossQ}. The resulting negative dispersion filter, when added to the gravitational wave interferometer and combined with frequency dependent squeezing, produces a sensitivity enhancement shown in Figure \ref{LIGOSens}(h). AlGaAs technology allows for resonators that give a 8 fold improvement in the sensitivity at 2 kHz detection frequency vs the Advanced LIGO quantum noise, given that the filter is maintained at or close to 4 K, so that $T\cdot Q_m^{-1} =$ 2$\times$ 10$^{-10}$. Due to the thermal expansion properties discussed in \ref{subsectionTHM}, it is possible to operate the filter cavity at temperatures up to 13 K while keeping the thermoelastic loss at acceptable level, however, this will degrade $T\cdot Q_m^{-1}$ to 6$\times$ 10$^{-10}$ if $Q_m$ is not increased. 

The performance depends critically on the mechanical and optical properties of AlGaAs/GaAs coatings. In order to fully realize the enhancement shown in figure \ref{LIGOSens}(h), we require an improvement in the test mass coating Brownian noise by a factor of more than two, which is theoretically possible using AlGaAs/GaAs coatings \cite{ColeCoherent}. Combination of WLSR with frequency dependent squeezing requires a filter cavity with minimal optical loss, which requires very low mode matching losses and optical coating losses. The size of the cat-flap and optical spring frequency are likely to be limited by the optical damage threshold of AlGaAs/GaAs coatings, for which we have used the lower limit of 4 MW/cm$^2$ discussed in section \ref{subsectionTHM}. Finally, we have assumed that a Q-factor $\sim$ 10$^{6}$ can eventually be achieved for an AlGaAs/GaAs cat-flap.The highest reported Q-factor of AlGaAs/GaAs at 4 K is 2 $\times$ 10$^{5}$. However, losses often depend on external factors such as clamping losses. It is also possible to reduce the loss of coatings by unconventional layer construction \cite{GorodetskyThermal, ColeCoherent}. Research into improved coatings is ongoing \cite{ColeCrystalline2, ColeSPIE}, and it is likely that better materials will enable even higher performing cat-flap resonators and test masses. Figure \ref{LIGOSens}(h) shows the effect of improving $T\cdot Q^{-1}_{\rm m}$ by a factor of 5, as well as increasing the optical spring frequency to 500 kHz. This leads to small but significant improvements in sensitivity and bandwidth, limited by quantum noise. The operating requirement for the WLSR filter does not utilize the highest possible value of $T\cdot Q^{-1}_{\rm m}$.

The $Q_m$ at the required $\omega_{opt}$ for the particular mirror size is acceleration loss limited when $\omega_0/(2\pi) =$ 250 Hz. Thus, extremely soft suspensions are not required to achieve the optimum Q-factor at 170 kHz. However, it is interesting to note that extremely high $Q_m$ at, for example, 10 kHz, can be achieved by using softer suspensions with $\omega_0/(2\pi) \sim$ 1 Hz, but this presents little advantage for the WLSR cavity. The fact that extremely delicate suspension ribbons are not required means that the resonators discussed here can be expected to be reasonably robust, having suspension frequency greater than 200 Hz. They should be relatively easily used in conjunction with a low vibration closed cycle refrigerator and integrated into a compact package.

An unstable optomechanical filter combined with frequency dependent squeezed vacuum can achieve broadband sensitivity enhancement of gravitational wave detectors. The unstable filter can be implemented using a crystalline AlGaAs/GaAs ``cat-flap'' resonator which is optically trapped within a DEMS cavity with \(T\cdot Q_m^{-1} = 2 \times10^{-10}\) and can enhance detector sensitivity to 2 kHz signals by a factor of 8. This will significantly improve the detection radius of less massive binary mergers and enable better exploration of the properties of neutron stars and black holes. 
In the future we will consider applications of WLSR in detuned interferometers and present results on the performance of the cat-flap resonators. Future work will also consider the design and testing of the feedback loop required for stabilization of the blue detuned filter cavity.

{\it Acknowledgements.}$-$We thank Haixing Miao, Garrett Cole, Markus Aspelmeyer, Shiuh Chao, Stefan Danilishin and Hamed Sadeghian for useful comments and collaborations. We would like to thank the LIGO Scientific Collaboration and its Optics Working Group for collaboration and reviewing this manuscript. This research was supported by the Australian Research Council (Grants No. DP17010442 and No. CE170100004).

\bibliography{WLrefs}
\end{document}